\documentclass[%
 preprint,
 superscriptaddress, 
 amsmath,amssymb,
 aps,
]{revtex4-1}

\usepackage{graphicx}
\usepackage{dcolumn}
\usepackage{bm}
\usepackage[draft=false]{hyperref}
\usepackage[mathlines]{lineno}
\usepackage{qcircuit} 
\usepackage{physics} 
\usepackage{subfig}  
\usepackage{tabularx}
\usepackage{threeparttable} 
\usepackage{tikz} 
\usepackage{pstricks}
\usepackage{pst-solides3d}
\usepackage{xcolor} 

\begin{document}

\title{Machine learning logical gates for quantum error correction}

\author{Hongxiang Chen}
\email{h.chen.17@ucl.ac.uk}
\affiliation{Dept. Computer Science, University College London}
\affiliation{Rahko Ltd., Finsbury Park,  N4 3JP, United Kingdom}
\author{Michael Vasmer}
\email{mvasmer@pitp.ca}
\affiliation{Perimeter Institute for Theoretical Physics, Waterloo, ON N2L 2Y5, Canada}
\altaffiliation[Also at ]{Institute for Quantum Computing, University of Waterloo, Waterloo, ON N2L 3G1, Canada}
\author{Nikolas P. Breuckmann}
\email{n.breuckmann@ucl.ac.uk}
\affiliation{Dept. of Physics \& Astronomy, University College London}
\author{Edward Grant}
\email{edward.grant.16@ucl.ac.uk}
\affiliation{Dept. Computer Science, University College London}
\affiliation{Rahko Ltd., Finsbury Park,  N4 3JP, United Kingdom}

\date{\today}

\begin{abstract}
Quantum error correcting codes protect quantum computation from errors caused by decoherence and other noise.
Here we study the problem of designing logical operations for quantum error correcting codes. 
We present an automated procedure which generates logical operations given known encoding and correcting procedures. 
Our technique is to use variational circuits for learning both the logical gates and the physical operations implementing them. 
This procedure can be implemented on near-term quantum computers via quantum process tomography. It enables automatic discovery of logical gates from analytically designed error correcting codes and can be extended to error correcting codes found by numerical optimizations.
We test the procedure by simulation on classical computers on small quantum codes of four qubits to fifteen qubits and show that it finds most logical gates known in the current literature. 
Additionally, it generates logical gates not found in the current literature for the [[5,1,2]] code, the [[6,3,2]] code, and the [[8,3,2]] code.
\end{abstract}

\maketitle


\section{\label{sec:intro}Introduction}

Quantum errors stem from undesired interactions with an outside environment.
There are many different interactions which may occur and their nature as well as their strength depend on the particular hardware architecture.
A natural assumption on the errors is that they are not adversarial, but are randomly distributed and localized.

It was shown by Shor~\cite{shor1995qec} that we can preserve quantum information by using \emph{quantum error correcting codes}.
The main idea is similar to what is done in classical error correction: redundancy is introduced by encoding some number of qubits $k$ into a larger number of \emph{physical qubits} $n$.
While for classical information (bits) we can introduce redundancy by keeping several copies, for quantum information it can be shown that copying of states is not possible by the laws of quantum mechanics, i.e. there is no device which takes any quantum state as an input and produces several copies of this state.
Despite this limitation it is possible to effectively encode a quantum state.
This is done by mapping the states of the $k$ logical qubits into non-local degrees of freedom of a highly entangled state of the physical qubits.
Although classical codes can not be used directly, there exists a construction due to Calderbank, Shor and Steane (CSS)~\cite{Calderbank1996,Steane1996} which takes two linear classical codes as an input and returns a quantum code.

In a quantum computer we manipulate quantum states using
a small set of unitary operators called quantum gates.
Gates applied to the encoded qubits are called \emph{logical gates}.
As the application of gates is prone to errors itself we would like to implement the logical gates using shallow depth circuits, meaning that only a small number of gates are being applied to the physical qubits to induce the logical gate.
Finding such circuits is a major challenge in quantum error correction and has thus far been done on a case-by-case basis.
For example, codes which are generated by the CSS construction (CSS codes) and encode a single logical qubit are known to have a fault-tolerant CNOT gate: it can be implemented on two copies of the same code by applying a CNOT between all pairs of physical qubits.


Here we apply the technique of variational circuit optimization to find fault tolerant logical gates for a given quantum error correcting code. Variational circuits have been used to parametrize wavefunctions, i.e. parametrize a unitary circuit transforming the $\ket{0}$ initial state. This approach has been used successfully to perform quantum chemistry calculations on Noisy Intermediate-Scale Quantum (NISQ) computers \citep{Peruzzo2014, Kandala2017, YingLiErrorMit, 1805.04492, 1808.10402}. The popular basic building blocks, called \textit{unit cells}, of these variational circuits are rotational gates $R_x,\,R_y,\,R_z$ and $\mathrm{CNOT}$ gates (see the Nomenclature and notation section of \cite{nelson_bible} for a definition). The specific mathematical properties of these gates makes optimizing the variational circuit relatively easy \cite{rotosolver, arXiv1903.12166}. In our case, we propose using variational circuits to find fault-tolerant logical gates for given quantum error correcting code.
Our procedure finds logical gates and their quantum circuit implementations by numerically optimizing
the variational circuit ansatz for both logical gates and their physical implementations (Fig.~\ref{fig:central_fig}).
Our procedure offers several benefits and much flexibility.
The ansatz used for the physical operation can be tailored to take advantage of the property of a specific quantum computing architecture.
The ansatz for the logical gate is variational and thus our procedure automates the discovery of logical gates given a quantum error correcting code.
The procedure can also target a specific logical gate when we fix the ansatz to this gate.
Therefore, for stabilizer codes and in particular for non-CSS codes, our procedure provides a straightforward first choice to find logical gates.
Furthermore, the procedure can be implemented on a quantum computer using quantum process tomography and is resource friendly for quantum codes requiring a small number of physical qubits.
It is hence feasible for an implementation on near-term quantum computers.

We note that in the literature, previous research has applied numerical optimization techniques to the field of quantum error correction for different purposes. Work in \citep{RobustQECViaCO, ChannelOptQEC, QECviaConvex, mario2018} used optimization algorithms (mostly convex optimization algorithms) to find error correcting quantum channels. Work in \citep{QVECTOR} demonstrated learning a circuit for preserving quantum information using a variational ansatz circuit. In \cite{CodesFromNN, RLforQEC} authors constructed quantum error correcting codes using neural networks. There is also a body of research using neural networks for decoding \cite{NeuralDecoderForTC, ScalableNNDecoder, RLDecoder, NNDecoderCircuitNoise, MLAssistedCor}. In this work, we applied similar optimization techniques to the novel problem of finding logical operators for quantum error correcting codes.

\textbf{Paper structure}. In this paper, we describe the procedure in detail in the Sec.~\ref{sec:results} 
where we also present the results of simulation on a classical computer where we apply the procedure to several CSS codes and non-CSS codes. 
The technical details of the simulation and the optimization are mentioned in Sec~\ref{sec:methods} and the data are attached with this paper in the Supplementary Information.
Among these results there are several new logical gates for the [[5,1,2]] code, the [[6,3,2]] code, and the [[8,3,2]] code, 
which to our knowledge have not appeared previously in the literature, and which we discuss in detail in the Sec.~\ref{sec:new_gates}. 
Finally, we make several comments on the benefits and disadvantages of our procedure in Sec~\ref{sec:discuss}.

\section{\label{sec:results}Results}

Here we present the procedure to find circuits which implement logical gates for error correcting codes. The procedure is inspired by the idea of circuit learning and uses ansatz circuits for both the logical gate and the physical operations that implement this logical gate in the encoded Hilbert space. Before we define this procedure, we first introduce the notation.

Commonly, an error correcting code encodes logical qubits (whose corresponding Hilbert space will be denoted as $H_A$),
into the subspace, denoted by~$L$, of another Hilbert space $H'=H_A\otimes H_B$. 
This mapping is unitary and is denoted as $E: H_A\to L \subset H'$. 
We call~$G$ a physical operation implementing the logical gate $g$ if it is
a unitary automorphism on~$L$ such that $E^{-1} G E \ket{\psi} = g \ket{\psi}$ 
for states $\ket{\psi}\in H_A$.

Now we present the procedure. We first describe how the procedure can be performed via simulation on a classical computer, and we describe its extension to quantum computer afterwards.
On a classical computer, we simulate the encoding, the physical operation, and the inverse encoding~$E^{-1}$. 
While the encoding/inverse encoding circuit is fixed by the choice of a particular error correcting code, we use ansatze for both the physical operation~$G$ and the logical gate~$g$. 
The ansatz for the physical operation is variational and may map logical states outside the logical space.
Because of this possibility, we instead apply a projector onto the codespace after the physical operation. This is achieved by measuring the stabilizer generators and applying a correction, where the correction is the minimum weight error compatible with the observed syndrome. 
The stabilizer measurements and corrections are simulated classically as a unitary circuit acting on the extended Hilbert space $H'\otimes H_C$. Here the qubits in the~$H_C$ are all initialized to the $\ket{0}$ state.
We use a variational circuit as the ansatz for the logical gate, to enable automatic discovery of logical gates for the quantum error correcting code. Alternatively, we can set a fixed unitary gate as the logical gate, and try to vary the ansatz for physical operation to find an implementation for this unitary gate.

\begin{figure}
\centering
\subfloat[Subfigure 1 list of figures text][Trial logical circuit]{
    \makebox[0.3\textwidth][c]{
    \Qcircuit @C=0.5em @R=1.0em @!R {
    \lstick{H_C: \ket{0}}     &\qw             &\qw                       &\qw                       &\multigate{2}{C}&\qw                     &\qw\\     
    \lstick{H_B: \ket{0}}     &\multigate{1}{E}&\multigate{1}{\mathcal{N}}&\multigate{1}{G(\theta_1)}&\ghost{C}       &\multigate{1}{E^\dagger}&\qw & \rstick{\ket{\phi^1_i}}\\ 
    \lstick{H_A: \ket{\psi_i}}&\ghost{E}       &\ghost{\mathcal{N}}       &\ghost{G(\theta_1)}       &\ghost{C}       &\ghost{E^\dagger}       &\qw \\
    {\gategroup{1}{6}{3}{6}{2.2em}{\}}}
    }
    }
\label{fig:subfig1}}
\qquad
\subfloat[Subfigure 2 list of figures text][Comparison circuit]{
    \raisebox{-53pt}{
        \makebox[0.3\textwidth][c]{
        \Qcircuit @C=0.5em @R=1.0em @!R {
        \push{\rule{1em}{0em}} & & \lstick{H_A: \ket{\psi_i}} & \gate{g(\theta_2)} & \qw & \rstick{\ket{\phi^2_i}}
        }}
    }
\label{fig:subfig2}}
\\
\subfloat[Subfigure 3 list of figures text][Procedure]{
    \includegraphics[width=0.8\textwidth]{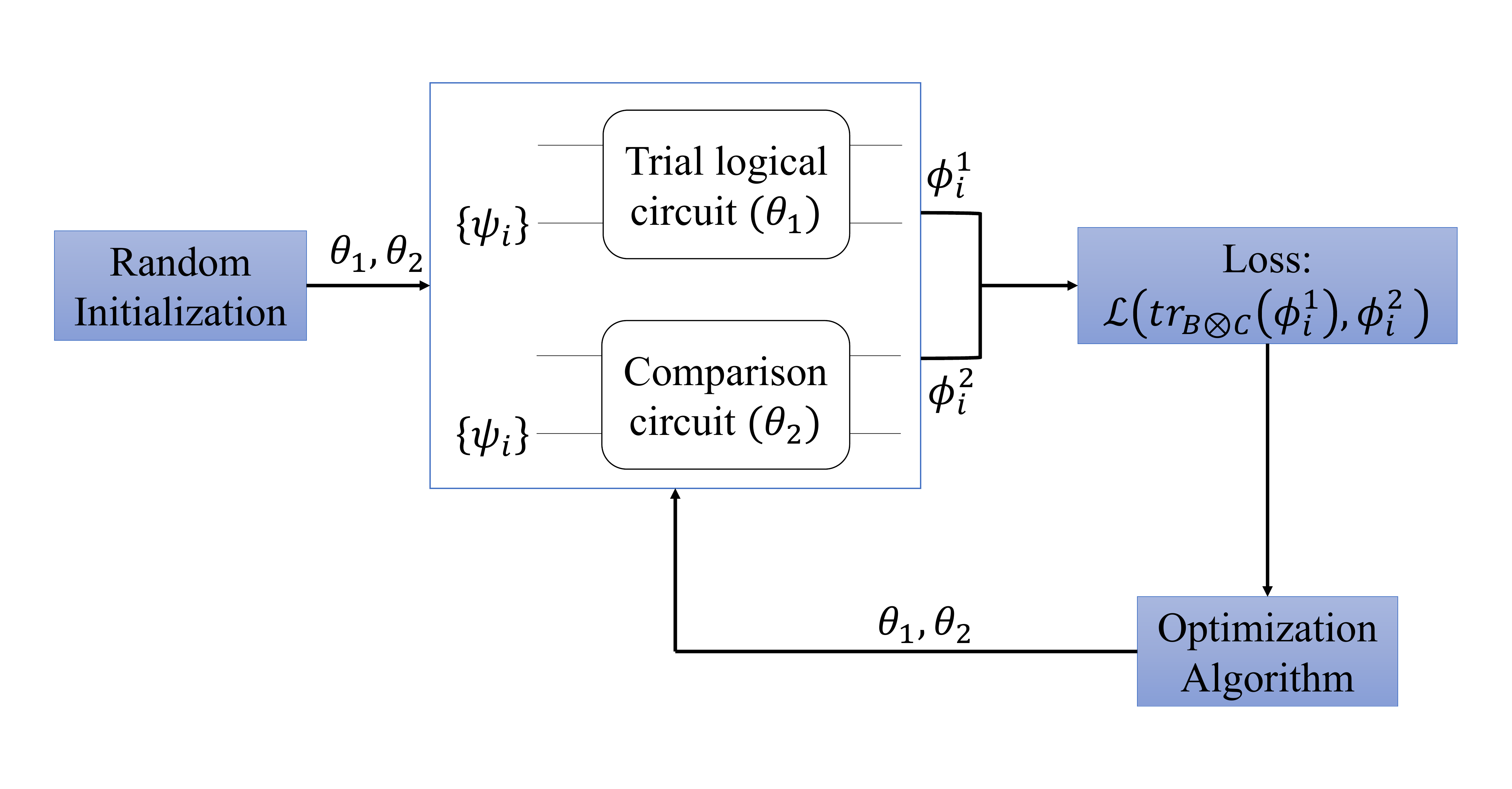}
\label{fig:subfig3}}
\caption{Learning logical gates \label{fig:central_fig}}
\begin{flushleft}
\footnotesize
To learn logical gates, we optimize angles~$\theta_1$ and~$\theta_2$ such that the trial logical circuit and the comparison circuit perform effectively the same unitary as measured by our loss function~$\mathcal{L}$.
Inside the trial logical circuit, the encoding/inverse encoding circuit~$E$/$E^\dagger$ maps input states $\{\psi_i\}_i^{N}$ into/out of the logical space. The ansatz $G(\theta_1)$ performs a series of physical operations and the circuit $C$ performs the stabilizer measurements and minimum weight error correction. In the comparison circuit, only the ansatz circuit $g(\theta_2)$ for the logical gate is performed.
The whole procedure starts with randomly initialize the two angles in both circuits, and then a tomographically complete set of quantum states are fed to the two circuits as inputs. 
Their outputs are gathered and fed to the loss function $\mathcal{L}$.
The calculated loss function values are fed to a minimization algorithm which outputs new angles for~$\theta_1$ and~$\theta_2$.
Then we rerun the trial logical circuit and the comparison circuit again and the whole loop continues until $\mathcal{L}$ is zero excluding float point errors, in which case the circuit $G(\theta_1)$ will perform the logical gate~$g(\theta_2)$ on the encoded space.
\end{flushleft}

\end{figure}

Our goal is to optimize the parameters for the ansatz circuits such that the for all possible input quantum states, a physical operation~$G$ together with the encoding, syndrome removal, and decoding circuit, act in the same way as a logical operation~$g$ (See Fig.\ref{fig:central_fig}). 
Specifically, we minimize the loss function

\begin{equation}\label{eq:loss_classical}
    \mathcal{L}(\theta_1, \theta_2) = \sum_{i} \left(
        1 - \mathrm{F}\left(
        \mathrm{tr}_{B\otimes C} \left( \ketbra{\phi_{1,i}} \right), \ketbra{\phi_{2,i}}
        \right) 
    \right) .
\end{equation}

Here $\phi^1_i$ ($\phi^2_i$) is the output of the trial logical circuit (comparison circuit) (see Fig.\ref{fig:central_fig}).
The input states $\{\psi_i\}_i^{N}$ form a tomographically complete set (the particular set of states we used in simulation is described in the Sec.~\ref{sec:tomography_complete}).
The trace $\mathrm{tr}$ is taken over the ancilla Hilbert space used for syndrome removal and the Hilbert space~$H_A$. 
For two quantum states~$\rho$ and~$\sigma$, we define~$\mathrm{F}(\rho,\sigma)$ to be the fidelity between them. When the minimization is successful and the average fidelity is zero excluding float point errors, the procedure succeeds in finding a logical gate for this code.

Our method can be adapted to run on a quantum computer in order to find logical gates for larger error correcting codes and to tailor the ansatz to the specific quantum computing architecture. We briefly outline such an extension here. The main change with regards to the classical implementation concerns the encoding and inverse encoding circuits. Instead of implementing a (non fault-tolerant) encoding circuit, we need a certain method of reliably preparing encoded Pauli eigenstates on the quantum computer. Given these states, we apply the logical operation ansatz in the same way as the classical implementation. We can also implement an error correcting procedure after applying the logical operation ansatz. Finally, instead of implementing the inverse encoding circuit, we envisage performing logical measurements of the encoded states in the $X$, $Y$ and $Z$ bases. Using these measurement outcomes, we perform logical state tomography~\cite{Vogel1989,nelson_bible} on the output states, and compare the tomography results with the unencoded states obtained via classical simulation, using an analogous loss function to the one shown in Eq.~\ref{eq:loss_classical}. 

The requirements of implementing our method on a quantum computer are relatively minor, as long as the number of logical qubits in the code is modest. 
The only subroutines we would need to implement on the quantum computer are: preparing encoded Pauli eigenstates, applying a physical operation ansatz, and measuring Pauli observables. 
Although the number of required logical Pauli measurements grow exponentially in the number of logical qubits, which might be a bottleneck of our proposed method. 
Potentially one can utilize a swap test\cite{PhysRevLett.87.167902, quant-ph0105032}  for calculation of the loss functions. 
Specifically, we may prepare in another error-corrected quantum computer the unencoded states, and analogously calculate the loss function in Eq.~\ref{eq:loss_classical} by a cross-device swap test.
We emphasize that our procedure can be applied to any code that can prepare encoded Pauli eigenstates and measure Pauli observables, i.e. it is not limited to qubit stabilizer codes.

\subsection{\label{sec:new_gates}Results for small codes}

We applied our procedure to a variety of small codes, as summarized in Table~\ref{tab:founded_gates_summary}. Of particular interest are the cases where non-Pauli logical gates were found by the experiment. In this section, we highlight the results of the experiment for the following codes: [[8,3,2]], [[8,2,2]], [[6,3,2]], and [[7,1,3]] surface code with a twist.

\newcommand{\pauligate}{Pauli group}
\newcommand{\newg}[1]{{\color{red}{\mathbf{#1}}}}
\begin{table}[]
\begin{threeparttable}
\caption{
Summary of quantum gates found by our procedure using classical simulation for a variety of small quantum error correcting codes.
}

\begin{tabularx}{1.0\textwidth}
{ | >{\raggedright\arraybackslash}X 
  | >{\raggedright\arraybackslash}X | }
\hline
{\bf Code}                                                                                                                                     & {\bf Logical gate found by the procedure}                             \\ \hline
[[4,1,2]]~\cite{Kesselring2018}, [[8,2,3]]~\cite{Grassl:codetables}, [[8,3,3]]~\cite{Gottesman1996,Steane1996a,Calderbank1997}, [[11,5,3]]~\cite{Steane1996a}, [[12,6,3]]~\cite{Grassl:codetables}, [[13,7,3]]~\cite{Grassl:codetables}, [[14,8,3]]~\cite{Grassl:codetables}, [[15,7,3]]~\cite{Calderbank1996,Steane1996a} & \pauligate                                                    \\ \hline
[[4,2,2]]~\cite{Grassl1997,Vaidman1996}                                                                                                                        & \pauligate, CNOT                    \\ \hline
[[5,1,2]]~\cite{Kubica2019}                                                                                                                        & \pauligate, $\newg{S}$, $\newg{H}$                    \\ \hline
[[5,1,3]] (Five-qubit code)~\cite{Laflamme1996,Bennett1996}                                                                                                                        & \pauligate, 
$e^{i\pi/4}SH$, 
$e^{i3\pi/4}XHXS^{\dagger}$, 
$e^{-i3\pi/4}XHXS$, 
$e^{-i\pi/4}HS^{\dagger}$, 
$e^{-i\pi/4}S^{\dagger}H$, 
$e^{-i3\pi/4}SXHX$, 
$e^{i3\pi/4}S^{\dagger}XHX$                      
\\ \hline
[[6,3,2]]~\cite{Grassl:codetables} & \pauligate, $\newg{H}_{12}\newg{CZ}_{12}\newg{H}_{12}$ \\ \hline
[[7,1,3]] (Steane code)~\cite{Steane1996,Calderbank1996}                                                                                                                        & \pauligate, Generators for the group generated by $H$ and $S$ \\ \hline
[[7,1,3]] (Surface code with a twist)~\cite{Yoder2017}                                                                                              & \pauligate, $e^{-i3\pi/4}SXHX$, $e^{i3\pi/4}S^{\dagger}XHX$   \\ \hline
[[8,2,2]] (Projective plane 2D color code)~\cite{Vuillot2019} & \pauligate, $CZ$, $H^{\otimes 2} CZ H^{\otimes 2}$, $H^{\otimes 2}SWAP$
\\ \hline

[[8,3,2]]~\cite{Kubica2015,Campbell2016}                                                                                                                        & \pauligate, $\newg{CZ}_{12}$, $\newg{CZ}_{13}$, $\newg{CZ}_{23}$, $CCZ$            \\ \hline
\end{tabularx}

\begin{tablenotes}
\item [1] The new logical gates we found, which have not been reported in the literature, are in highlighted in red and in boldface. For these logical gates, the parity check matrix of the corresponding quantum code and the physical operations which implement them are provided in OpenQASM\citep{openqasm} format in the Supplementary Information.
\item [2] The exact experimental configuration and optimization algorithm we used is discussed in the Sec.\ref{sec:methods}.
\item [3] We found a generating set of logical Pauli gates for the codes which we have labeled Pauli group in the column \textit{Logical gate found by the procedure}.
\end{tablenotes}
\label{tab:founded_gates_summary}
\end{threeparttable}

\end{table}

The [[8,3,2]] code is the smallest non-trivial 3D color code~\cite{Kubica2015,Campbell2016}. We found three transversal Clifford gates for this code which (to the best of our knowledge) have not previously appeared in the literature: $CZ_{12}$, $CZ_{13}$, and $CZ_{13}$, where $CZ_{ij}$ denotes a logical~$CZ$ acting on logical qubits $i$ and $j$. Our optimization procedure also provided a simple implementation of these gates, which we now explain. 

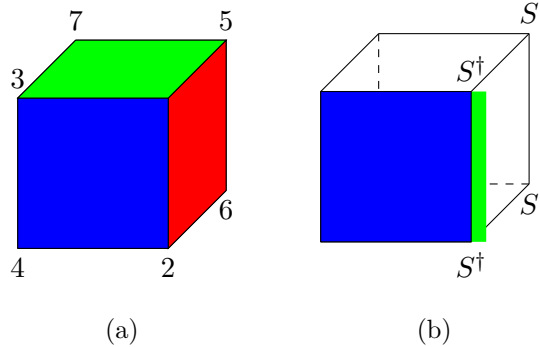
\begin{figure}[h]
    \centering
    \subfloat[]{
        \begin{tikzpicture}[scale=0.5]
            \coordinate[label=below:8] (A) at (0,0,0);
            \coordinate[label=below:4] (B) at (0,0,4);
            \coordinate[label=above:7] (C) at (0,4,0);
            \coordinate[label=above:3] (D) at (0,4,4);
            \coordinate[label=below:6] (E) at (4,0,0);
            \coordinate[label=below:2] (F) at (4,0,4);
            \coordinate[label=above:5] (G) at (4,4,0);
            \coordinate[label=above:1] (H) at (4,4,4);
            \coordinate[label={[label distance=-0.5cm]0:$\overline{X}_{1}$}] (X1) at (4.25,2,2);
            \coordinate[label={[label distance=-0.5cm]0:$\overline{X}_{2}$}] (X2) at (2.5,2.1,4);
            \coordinate[label={[label distance=-0.5cm]0:$\overline{X}_{3}$}] (X3) at (2,4,2);
            \draw [dashed] (A) -- (B);
            \draw [dashed] (A) -- (C);
            \draw [dashed] (A) -- (E);
            \draw (B) -- (D);
            \draw (B) -- (F);
            \draw (C) -- (D);
            \draw (C) -- (G);
            \draw (D) -- (H);
            \draw (E) -- (F);
            \draw (E) -- (G);
            \draw (F) -- (H);
            \draw (G) -- (H);
            \draw[fill=red, fill opacity=0.2] (E) -- (F) -- (H) -- (G) -- cycle;
            \draw[fill=blue, fill opacity=0.2] (B) -- (F) -- (H) -- (D) -- cycle;
            \draw[fill=green, fill opacity=0.2] (C) -- (D) -- (H) -- (G) -- cycle;
        \end{tikzpicture}
    }%
    \qquad
    \subfloat[]{
        \begin{tikzpicture}[scale=0.5]
            \coordinate (A) at (0,0,0);
            \coordinate (B) at (0,0,4);
            \coordinate (C) at (0,4,0);
            \coordinate (D) at (0,4,4);
            \coordinate[label=below:$S$] (E) at (4,0,0);
            \coordinate[label=below:$S^{\dagger}$] (F) at (4,0,4);
            \coordinate[label=above:$S$] (G) at (4,4,0);
            \coordinate[label=above:$S^{\dagger}$] (H) at (4,4,4);
            \draw [dashed] (A) -- (B);
            \draw [dashed] (A) -- (C);
            \draw [dashed] (A) -- (E);
            \draw (B) -- (D);
            \draw (B) -- (F);
            \draw (C) -- (D);
            \draw (C) -- (G);
            \draw (D) -- (H);
            \draw (E) -- (F);
            \draw (E) -- (G);
            \draw[green, line width=0.4cm] (F) -- (H);
            \draw (G) -- (H);
            \draw[fill=blue, fill opacity=0.2] (B) -- (F) -- (H) -- (D) -- cycle;
        \end{tikzpicture}
    }
    \caption{$CZ$ gates in the [[8,3,2]] code. Qubits are placed on the vertices of the cube. In (a), we highlight three logical $X$ operators of the code, namely $\overline{X}_{1}=X_{1}X_{2}X_{5}X_{6}$ (red face), $\overline{X}_{2}=X_{1}X_{2}X_{3}X_{4}$ (blue face), and $\overline{X}_{3}=X_{1}X_{3}X_{5}X_{7}$ (green face). Three corresponding logical $Z$ operators are: $\overline{Z}_{1}=Z_{1}Z_{3}$, $\overline{Z}_{2}=Z_{1}Z_{5}$, and $\overline{Z}_{3}=Z_{1}Z_{2}$. (b) Suppose that we apply $S$ and $S^{\dagger}$ gates to the vertices on the face that supports $\overline{X}_{1}$. This unitary maps $\overline{X}_{2}$ (blue face) to $\overline{X}_{2}\overline{Z}_{3}$ (blue face and green edge). Likewise, $\overline{X}_{3}$ is mapped to $\overline{X}_{3}\overline{Z}_{2}$. Therefore, this unitary implements a logical $\overline{CZ}_{23}$ gate.}
    \label{fig:CZ_832}
\end{figure}

Logical gates in the [[8,3,2]] code can be understood geometrically. Suppose we place qubits on the vertices of a cube, as shown in Figure~\ref{fig:CZ_832}. The stabilizer group of the [[8,3,2]] code can be generated by an operator consisting of Pauli-$X$ operators acting on every qubit, alongside operators that are associated with the faces of the cube, i.e.\ for each face, $f$, we have an operator $\Pi_{v\in f}Z_{v}$, where $Z_{v}$ denotes a Pauli-$Z$ operator acting on the qubit on vertex~$v$. With this definition, the logical $X$ operators of the code are associated with the faces of the cube ($\overline{X}_{1}=\Pi_{v\in f}X_{v}$ etc.) and the logical $Z$ operators are associated with the edges of the cube. We note that opposite faces support logical $X$ operators that act on the same encoded qubit, and the corresponding logical $Z$ operators are supported on the edges linking these faces. 

To implement a logical $\overline{CZ}_{ij}$ gate, we apply $S=\text{diag}(1,i)$ and $S^{\dagger}$ gates in an alternating pattern to the vertices of a face that supports $\overline{X}_{k}$. We show this operator in Figure~\ref{fig:CZ_832}b and we denote it by $U$. We now show that $U$ implements a $\overline{CZ}_{ij}$ gate: namely it maps $\overline{X}_{i}$ to $\overline{X}_{i}\overline{Z}_{j}$, and has no effect on all other logical operators. First, we note that all operators that consist exclusively of Pauli-$Z$ operators are unaffected by $U$ because $S$ and $Z$ commute. Therefore, $Z$-type stabilizers and logical $Z$ operators are mapped to themselves by $U$. Next we consider operators that contain Pauli-$X$, which are transformed by $S$ as follows: $SXS^{\dagger}=Y=iXZ$. Consider the $X$ stabilizer of the code ($X$ on all the qubits). It is straightforward to see that~$U$ maps this operator to a product of itself and a $Z$ stabilizer associated with the face where we applied $U$ (the alternating pattern causes the factors of $i$ and $-i$ to cancel). Finally, consider $X_{i}$ (the blue face in Figure~\ref{fig:CZ_832}b). This operator is mapped to a product of itself and a $\overline{Z}_{j}$ operator with support on an edge that links the $\overline{X}_{j}$ faces (the green edge in Figure~\ref{fig:CZ_832}b). Similarly, $\overline{X}_{j}$ is mapped to $\overline{X}_{j}\overline{Z}_{i}$

In addition, our procedure found a transversal implementation of $CCZ$ for the [[8,3,2]] code. This gate was known previously~\cite{Kubica2015,Campbell2016}, however the fact that our procedure found a non-Clifford gate ($CCZ$) with no prior knowledge about the structure of the code is notable. It is often relatively straightforward to implement Clifford gates and Pauli measurements in error correcting codes. Such operations are classically simulable, but they can be promoted to universality by adding a single non-Clifford gate. However, codes with transversal non-Clifford gates are rare, and understanding the structure of such codes is an active area of research~\cite{Kubica2015a,Kubica2015,Vasmer2019,Vuillot2019,Rengaswamy2019}. Most examples of constructions of code families with non-Clifford gates exploit some specific structure of the code family, such as tri-orthogonality~\cite{Bravyi2012,Campbell2017,Campbell2017a}. Our procedure may be capable of finding fault-tolerant non-Clifford gates for codes whose structure is opaque to us. In particular, the structure of non-CSS codes with non-Clifford gates is poorly understood, so our procedure may be able to shed some light on this area given enhanced computational resources.

We successfully found non-Pauli gates for codes related to 2D color codes: the [[8,2,2]] code (2D color code defined on a projective plane~\cite{Vuillot2019}) and the [[6,3,2]] code (subcode of a 2D color code defined on a hexagon~\cite{Criger2016}). For the [[8,2,2]] code, we found the gates that we would expect to find in a 2D color code with two logical qubits: logical $CZ$ implemented by transversal $S$ and $S^{\dagger}$, logical $H^{\otimes 2} CZ H^{\otimes 2}$ implemented by transversal $\sqrt{X}$ and logical $H^{\otimes 2}SWAP$ implemented by transversal $H$. And we found that the [[6,3,2]] code inherits one of the transversal gates of its parent 2D color code (the [[6,4,2]] code~\cite{Criger2016}): $CZ_{12}$ implemented by transversal $S$ and $S^{\dagger}$.

\begin{figure}[ht]
    \centering
    \includegraphics{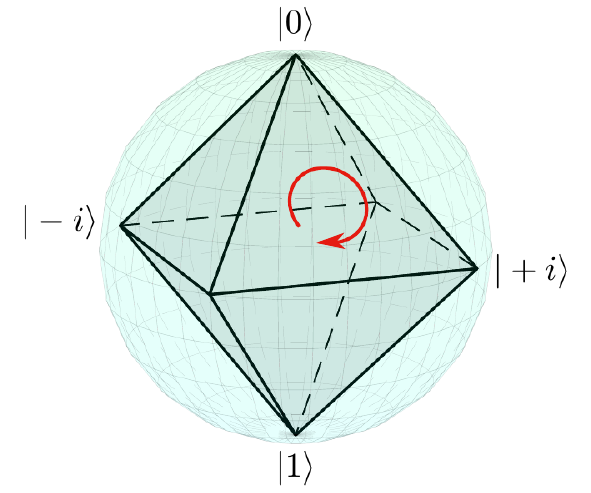}
    \caption{An illustration of the Clifford gate $K_{1,1,1}=e^{i\pi/4}SH$. This gate maps $\ket{0}\rightarrow\ket{+i}$, $\ket{+i}\rightarrow\ket{+}$, and $\ket{+}\rightarrow\ket{0}$ (up to global phases). It can be understood geometrically by considering an octahedron embedded in the Bloch sphere. The operation performed by the gate is a clockwise rotation of $2\pi/3$ around the axis which is normal to the face marked with a red circle. The $\ket{+}$ state is at the front unlabeled vertex of the octahedron.}
    \label{fig:my_label}
\end{figure}

Our procedure also found transversal Clifford gates for the [[7,1,3]] surface code with a twist. As was noted in~\cite{Yoder2017}, this code has transversal implementations of the octahedral Clifford gates, which are defined as follows:
\begin{equation}
    K_{x,y,z}=\exp{i\frac{\pi}{3\sqrt{3}}(xX+yY+zZ)},
\end{equation}
where $x,y,z\in{-1,1}$. These gates cyclically permute the Pauli operators. We can decompose these octahedral gates into products of more familiar Clifford gates as follows:
\begin{equation}
\begin{split}
    &K_{1,1,1}=e^{i\pi/4}SH, \\
    &K_{1,-1,1}=e^{i\pi/4}HS, \\
    &K_{1,1,-1}=e^{i3\pi/4}XHXS^{\dagger}, \\
    &K_{-1,1,1}=e^{-i3\pi/4}XHXS, 
\end{split}
\qquad
\begin{split}
    &K_{-1,-1,-1}=e^{-i\pi/4}HS^{\dagger}, \\
    &K_{-1,1,-1}=e^{-i\pi/4}S^{\dagger}H, \\
    &K_{-1,-1,1}=e^{-i3\pi/4}SXHX, \\
    &K_{1,-1,-1}=e^{i3\pi/4}S^{\dagger}XHX.
\end{split}
\end{equation}
The specific gates we found in our experiments were transversal realizations of $K_{-1,-1,1}$ and $K_{1,-1,-1}$.

\section{\label{sec:discuss}Discussion}

\newcommand{\rotosolve}{\texttt{Rotosolve}}

We comment on several aspects of our procedure here.
Firstly, we mention here a rough resource estimate of our proposed procedure. For the calculation of the loss function in Eq.~\ref{eq:loss_classical}, we require the the full process tomography of the encoded logical space, which requires a number of quantum circuit runs growing exponentially with respect to the number $k$ of logical qubits. However, this cost is benign when $k$ is small and can further mitigated by running these circuits in parallel on multiple computers. Also, due to the fact that transversal ansatz is mostly likely to be used as the ansatz for logical gates, we expect a friendly linear growth of the number of parameters with respect to the number of physical qubits ($n$) in Eq.\ref{eq:loss_classical} which we need to optimize. In the classical simulations we have done, typically one optimization converge in less than 5 hours running on a commercially available GPU (Nvidia~RTX~2080~Ti).

Secondly, our optimization benefits from the use of \rotosolve \citep{rotosolver} optimization algorithm, which depends largely on the fact that the functional dependency of the lost function in Eq.\ref{eq:loss_classical} is sinusoidal due to the parametrized rotation gates we have used for the ansatze (see Appendix~\ref{sec:expsetup}).
Adapting our procedure for specific quantum computation hardware might require a different set of parametrized gates and might invalidate the use of \rotosolve. In this case, the optimization will be harder due to the non-convex nature of the loss function and the high number of parameters.

A promising future research avenue would be to use our procedure to explore quantum codes that have not been as extensively studied as qubit stabilizer codes. Qudit quantum codes are a natural example, where a qudit is the $d$-dimensional analogue of a qubit. The stabilizer formalism can be extended to prime (or prime power) qudits~\cite{Gottesman1999}, which means that it would be straightforward to generalize our procedure to these cases. Considerably less research has been done into implementing logical gates in qudit stabilizer codes compared with qubit stabilizer codes, so we may be able to find more unknown fault-tolerant gates in the qudit context. In addition, we emphasize that our procedure is not limited to stabilizer codes, and can be applied to non-stabilizer codes e.g.\ the codes described in~\cite{Cross2009}.

In summary, we have proposed a procedure to automate the discovery of logical gates using shallow quantum circuits for a given quantum error correcting code.
The ansatz for the logical gate can be tailored to a specific quantum computing architecture to take advantage of this architecture.
We have shown that it can find logical gates available in the current literature for a number of error correcting codes and it additional produces new logical gates for the [[5,1,2]] code, the [[6,3,2]] code, and the [[8,3,2]] code.
Although the procedure is simulated classically, we have proposed an extension of this procedure on quantum computers and we believe an implementation on near-term quantum computers for error correcting codes requiring a small number of qubits is feasible.

\section{Methods\label{sec:methods}}

\subsection{Experimental setup for Table~\ref{tab:founded_gates_summary}\label{sec:expsetup}}

In the simulation experiments shown on Table~\ref{tab:founded_gates_summary} (except for the [[5,1,2]] code), the ansatz for logical operation is transversal, which is naturally fault-tolerant. A transversal ansatz is formed by using three single qubit rotation gates ($\mathrm{R_{j}} = e^{-i\theta \sigma_{j}/2}$, where $\{\sigma_{j}\}_{j=x, y, z}$ are the three Pauli matrices) on each physical qubit, where each rotation gate has its own angle that can be adjusted independently of other rotation gates.
For the parameterization of the logical gate $g(\theta_2)$, we use the ansatz which parametrizes arbitrary unitary transformation on the Hilbert space $H_A$. 
Denote the number of qubits in $H_A$ as $n_a$. 
When $n_a=1$, the ansatz is simply the three single qubit rotation gates mentioned before.
When $n_a=2$, the ansatz is the circuit shown in Fig.2 in \cite{quant-ph-0308033}.
When $n_a=3$, we obtain a circuit parametrization for arbitrary three qubit unitary gates using the \texttt{QSD} decomposition provided by \citep{universalQCompiler}.
When $n_a > 3$, only the first three qubits are selected on which we apply the three qubit ansatz mentioned previously. This is because the exponential increase of the possible ansatze makes experimentation infeasible.

We note that for the experiments on the code $[[4,2,2]]$ in the table, the encoder used only encodes one of the two logical qubits and we only experimented on this logical qubits. 
In addition, as an early stage proof of principle that our procedure can find entangling gates, we targeted the logical gate CNOT for this code $[[4,2,2]]$ in one experiment
In this experiment, we encoded a pair of qubits in two copies of the [[4,2,2]] code (one encoded qubit per code). We then used a transversal two-qubit gate ansatz, where each gate coupled corresponding qubits in the different codes. Using this ansatz, we were able to find a transversal logical CNOT gate between the codes. The physical operation that implemented this logical gate is simply CNOT gates between corresponding qubits in the different codes. This shows that our method is capable of finding gates acting between separate codes.

In the case of the [[5,1,2]] code, we used a non-trasversal but still fault-tolerant ansatz for the logical gate. Specifically, we used a transversal ansatz (as described above) for three of the qubits, and an parametrization of an arbitrary two-qubit unitary for the final two qubits. This choice of ansatz was motivated by the structure of the code~\cite{Kubica2019}. The fact that we found logical gates (including non-Pauli gates, see Table~\ref{tab:founded_gates_summary}) using this ansatz shows that our procedure can work well with ans\"{a}tze that are tailored to a particular code.

For minimization, we used the minimization algorithm \rotosolve\cite{rotosolver} to minimize the $\mathcal{L}$. 
\texttt{Rotosolve} specializes in minimizing function of the form $\bra{0} U(\theta)^\dagger H U(\theta) \ket{0}$, where $U$ is the unitary transformation of a quantum circuit made from constant unitary gates and variational rotation gates $\mathrm{Rx}$, $\mathrm{Ry}$, and $\mathrm{Rz}$. 
It is easy to see our loss function defined in Eq.~$\ref{eq:loss_classical}$ follows the same form if we consider $\Tr_\mathrm{ancilla}(\ketbra{\phi_{2,i}})$ to be the $H$, and $g(\theta_2)$ to be the $U$.

We note that for performance issues, the simulation was carried out using proprietary software written by the author HX for Rahko Ltd.

\subsection{Tomographically Complete Set\label{sec:tomography_complete}}

For a quantum channel $\varepsilon$, its action on the any quantum state $\rho$ can be uniquely determined by 
its action on some $\{\psi_i\}$ which forms a tomographically complete set\citep{nelson_bible}.
An straight-forward example of a tomographically complete set is $\{\psi_i\}$ such that 
$\{\ketbra{\psi_i}\}$ forms a basis for all density matrices. In our case, we only need a set of $1$ qubit states such that $\{\ketbra{\psi_i}\}$ forms a basis for all $2\times 2$ density matrices, since for $2^n\times 2^n$ density matrices, the tensor products of this set of $1$ qubit states form a complete tomography set.

Now we describe the six states, which we call the \textit{six Bloch states}, which we use as the initial states for our experiments. These states are,

\begin{align} 
    \{\ketbra{\psi_i}\} = \{
    \ket{0}, \ket{1}, \ket{+}, \ket{-}, \ket{+i}, \ket{-i}
    \}
\end{align}

Here we show that any $2\times 2$ matrix can be expressed (non-uniquely) by the density matrices corresponding to the six Bloch states. 
And therefore, any $n$-qubit density matrix can also be expressed by tensor products of the six Bloch states. 
Therefore, tensor products of the 
six Bloch states forms a complete tomography set for $n$-qubit quantum channel.

Let $\vec{a}=(a,b,c,d,e,f\}$, and let $\vec{b'}$ be the density matrices of states in $\{\ketbra{\psi_i}\}$. We can check that,

$$\vec{a} \vdot \vec{b'} = 
\left(
\begin{array}{cc}
 a+\frac{c}{2}+\frac{d}{2}+\frac{e}{2}+\frac{f}{2} & \frac{c}{2}+\frac{f i}{2}-\frac{d}{2}-\frac{i e}{2} \\
 \frac{c}{2}+\frac{e i}{2}-\frac{d}{2}-\frac{i f}{2} & b+\frac{c}{2}+\frac{d}{2}+\frac{e}{2}+\frac{f}{2} \\
\end{array}
\right).
$$

To express any $2 \times 2$ matrix $\rho$ as linear combinations of $\vec{b'}$, we consider the 
linear equation $\vec{a}\vdot \vec{b'} = \rho$. It can be written as $A\vdot \vec{a} = \vec{c}$,
where $\vec{c}$ is the vector of elements of $\rho$, $A$ is the coefficient matrix:
$$
A = \left(
\begin{array}{cccccc}
 1 & 0 & \frac{1}{2} & \frac{1}{2} & \frac{1}{2} & \frac{1}{2} \\
 0 & 0 & \frac{1}{2} & -\frac{1}{2} & -\frac{i}{2} & \frac{i}{2} \\
 0 & 0 & \frac{1}{2} & -\frac{1}{2} & \frac{i}{2} & -\frac{i}{2} \\
 0 & 1 & \frac{1}{2} & \frac{1}{2} & \frac{1}{2} & \frac{1}{2} \\
\end{array}
\right).
$$

Since the rank of $A$ is $4$, the rank of the augmented matrix $[A|c]$ is also $4$, hence the above linear equation $A\vdot \vec{a} = \vec{c}$ has infinitely many solutions.

\bibliography{references.bib}

\begin{acknowledgments}
We wish to acknowledge the usage of high performance computing cluster from Department of Computer Science, University College London in completion of this project. H.C. acknowledges the support though a Teaching Fellowship from UCL. Research at the Perimeter Institute is supported in part by the Government of Canada through the Department of Innovation, Science and Economic Development Canada and by the Province of Ontario through the Ministry of Economic Development, Job Creation and Trade. N.P.B. is supported by the UCLQ Fellowship. E.G. is supported by the UK Engineering and Physical Sciences Research Council (EPSRC) [EP/P510270/1]
\end{acknowledgments}

\end{document}